# Narrative to Trajectory (N2T+): Extracting Routes of Life or Death from Human Trafficking Text Corpora

Saydeh N. Karabatis
Vandana P. Janeja
{saydeh1,vjaneja}@umbc.edu
University of Maryland, Baltimore County (UMBC) Baltimore,
Maryland, USA

## ABSTRACT

Climate change and political unrest in certain regions of the world are imposing extreme hardship on many communities and are forcing millions of vulnerable populations to abandon their homelands and seek refuge in safer lands. As international laws are not fully set to deal with the migration crisis, people are relying on networks of exploiting smugglers to escape the devastation in order to live in stability. During the smuggling journey, migrants can become victims of human trafficking if they fail to pay the smuggler and may be forced into coerced labor. Government agencies and anti-trafficking organizations try to identify the trafficking routes based on stories of survivors in order to gain knowledge and help prevent such crimes. In this paper, we propose a system called Narrative to Trajectory (N2T$^+$), which extracts trajectories of trafficking routes. N2T$^+$ uses Data Science and Natural Language Processing techniques to analyze trafficking narratives, automatically extract relevant location names, disambiguate possible name ambiguities, and plot the trafficking route on a map. In a comparative evaluation we show that the proposed multi-dimensional approach offers significantly higher geolocation detection than other state of the art techniques.

## CCS CONCEPTS

• **Information systems** → **Information integration**; • **Computing methodologies** → *Natural language processing*; • **Applied computing** → Anthropology.

## KEYWORDS

Data Science, Natural Language Processing, Human Trafficking, Location Name Disambiguation

## 1. INTRODUCTION

Both climate change and political unrest can lead to lack of resources, poverty, and instability that can force certain populations of affected areas to depart from their homeland and seek refuge in far away places. Since they lack proper documentation to legally enter non-neighboring countries, these migrants rely on a network of smugglers to facilitate their journey during which many human rights are violated. In addition, the smuggling process is a highly profitable money laundering business in which the migrant pays an exorbitant amount of money to the traffickers in order to be secretly moved from one location to another [12].

The Department of Economics and Social Affairs in the United Nations (UN) created a list of 17 Sustainable Development Goals (SDG) for peace and prosperity for all people. These goals provide an urgent call for action by all UN member countries that specifically target sustainability and development for underdeveloped countries [11]. Three out of the 17 SDGs target human trafficking (HT) thus making it easier for anti-trafficking organizations to address this "grave human rights violation" [11]. SDG 5 aims to "achieve gender equality and empower all women and girls" (forced sex exploitation, a form of HT violates this goal). SDG 8 aims to promote inclusive economic growth and "decent work for all" (unpaid/underpaid labor, a form of HT violates this goal). SDG 16 aims to "promote peaceful and inclusive societies" and "provide access to justice for all" (migrant smuggling across different countries violates this goal).

Climate change has been setting serious obstacles to various communities in Far East Asia and in the Pacific region. "This region includes 13 out of 30 countries most vulnerable to the impacts of climate change" [14]. The impacted communities have been seeing more people fall into poverty and seeking refuge outside their homeland. In addition, the unrest in some of the Near East countries, mainly Syria and Lebanon, has led to the displacement of several million people who have been looking for refuge in safer non-neighboring countries. In both cases, the migrants often rely on a network of criminal smugglers to facilitate their movement and to assist them during their journey. Those who succeed in reaching their destination must deal with the trauma experienced throughout their journey, however, not everyone is successful as there have been several reports of organ harvesting from migrants who fail to pay the expenses associated with the trip or migrants losing their lives along the journey before reaching safe haven [5].

It is commonly known that most migrants take the same paths that others have crossed earlier. These paths are often referred to as routes of life or death [13]. One way law enforcement agencies can disrupt migrant trafficking activities and save the lives of the migrants is to predict the route by gaining knowledge about past trafficking routes. Non-profit organizations can reach out to victims during their trafficking routes so that they do not have to endure the abuse at the hands of their traffickers.

Many narratives posted on media and anti-trafficking organizations sites contain a varying number of details about the experience of survivors during their captivity. Those narratives are either written by journalists or by members of the humanitarian organizations following interviews with the survivors. In order to protect the identity of the victims, most media sites often provide vague location names of the route travelled by the victims. Very few narratives tell it all: place of origin, visited places along the transportation routes, and the destination. We were able to locate articles that narrate stories of victims and list the location names along the transportation route in chronological order without disclosing any personal information of the victims.

Given the overwhelming number of survivor narratives reported on the internet and since very few of these narratives contain particular names of places along the transporting routes, it is not feasible for a human to read each narrative, extract and identify the location names, and plot on a map the travelled trajectory. Therefore, it is necessary to create a tool that automatically parses the narratives, extracts location names, disambiguates these names if ambiguation is present, and plots the transportation route as narrated on a map. Such a tool helps the officials identify previous trajectories and gain insight of future routes in order to save the lives of the vulnerable trafficked victims.

The complexity of human trafficking activities requires both human and machine intelligence to untangle. Using web advertisement as input, Szekely et al. apply Artificial Intelligence (AI) methods to extract telephone numbers of sex laborers, Esfahani et al. propose a semi-automatic composite model to identify sex trafficking ads, Tong et al. describe the development of a multimodal deep learning model to identify sex trafficking ads, and Nagpal et al. propose an entity resolution pipeline to extract clusters of data with prior history of human trafficking activities [7–10]. None of these works mines location names from the advertisements. Extracting geographical tags from atypical language models posted on illicit activities websites is presented in [2] using DARPA MEMEX [4]. The authors describe an Integer Linear Programming (ILP) model that processes human trafficking advertisements and produces a set of geolocation names. This approach works only if the population of the extracted location names exceeds 15 thousand. It does not address ambiguity of location names. Molina-Villegas et al. describe a geographic name entity recognition and disambiguation model *GNER* in Mexican news articles using word embedding and semantics to develop a Mexican Geoparser[6]. The model achieves acceptable accuracy in recognizing geographic named entity, but fails to fully resolve location name ambiguity.

Most of the above referenced works do not mine geographic locations and the few that do have major limitations. We proposed a Narrative to Trajectory (N2T) prototype system that processes narratives to identify trajectories [3]. The initial version of N2T lacks disambiguating location names. In this paper we propose N2T[+], a Data Science (DS) and AI model for disambiguating location names and tracing precise human trafficking trajectories extracted from text corpora of victim narratives. The contributions to this paper are to:

- Identify location names from text corpora
- Disambiguate location names in the presence of ambiguities
- Allocate the precise coordinates to each location name

The remainder of the paper is as follows: In Section 2, we describe our novel methodology to identify the victim's trajectory. In Section 3, we evaluate our proposed prototype through experiments. In Section 4 we conclude our paper and propose future work of the N2T[+] system.

## 2. METHODOLOGY

N2T[+] accepts a narrative as input, preprocesses it, splits it into tokens, labels each token according to its semantics and its syntax in the sentence, disambiguates location tokens when ambiguity is present, assigns precise spatial coordinates to location tokens, and outputs the trajectory as narrated on a map. To do so, N2T[+] integrates structured (Gazetteer and Lexicon) with unstructured (text corpora) heterogeneous data, performs tokenization methods, and leverages contextual sliding widow and rules. A high level architecture of the N2T[+] system is illustrated in Fig. 1.

N2T[+] consists of five components. The early version of N2T[+] does not address location name disambiguation [3]. For this reason, we use a contextual sliding window, rules, and DB techniques to enhance the early version and resolve location name ambiguity.

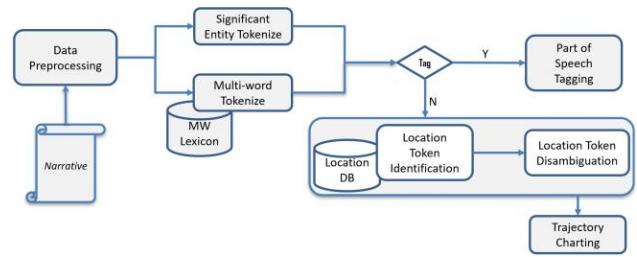

**Figure 1: N2T[+] System Architecture**

**Data Preprocessing**: Transforms the input text into a format applicable for analysis and prediction by performing data cleansing and replacing special and non-ASCII characters with NLP acceptable characters.

**Tokenization**: Splits text into meaningful tokens $To\_V$ by using NLP techniques that identify the divider between words in the text and generate tokens composed of either single or multi-words. However, the concept of using dividers (e.g. blank) by themselves for the purpose of generating tokens has its own limitations because some tokens are made out of multi-words where the designated divider can also be a unifier as in 'New Orleans'. For this purpose, we use:

- Significant Entity Tokenization (ST): It segments text into potential tokens of interest by applying lexical semantic rules specific to the language of the text. The output of this method is an ordered list of entities $To\_V_S = [(to\_v_i)]$ that are considered significant to the tokenizer.

- **Multi-word Tokenization (MWT)**: It first applies sentence okenization to the input. Each divided sentence is then tokenized into single tokens. Augmenting the generated list of tokens with the MW Lexicon helps generate tokens that also include multiple words. Lexicon helps generate tokens that also include multiple words. The output of this method is an ordered list of single word or multi-word tokens $To\_V_M = [(to\_v_j)]$. At the end of the Tokenization step, $To\_V = To\_V_S \vee To\_V_M$.

**Geospatial Token Identification and Disambiguation**: *Tokenization* transforms the unstructured data into an ordered list of tokens $To\_V$ for each narrative. However, not all the identified tokens signify locations. In this step, we identify and disambiguate all the location tokens and assign the precise coordinates using a contextual sliding window, rules, and integration methods.

*Location Disambiguation*: It is commonly known that not all location names are unique. It is possible to encounter the same name that refers to different geographic places in the world: the city Tripoli is found in Greece, Lebanon, and Libya. We refer to this situation as *Homonym Ambiguity*. In addition, not all locations carry a distinct name. For example, Bombay is formally known as Mumbai. We refer to this situation as *Synonym Ambiguity*. To resolve the location ambiguity, we try to find the exact country/state where this location belongs to. Identifying the country/state helps in specifying the geospatial coordinates of the location token.

Our N2T[+] system includes a lookup Location dimension *Loc* that contains location names along with their country/state, longitude, and latitude values [*(to_v, to_cntry, to_long, to_lat)*] obtained from GeoNames [1]. We update *Loc* and add to it two fields [*(to_ha, to_sa)*] that signify if an entry is of *Homonym Ambiguity* or *Synonym Ambiguity* type respectively. The *Homonym Ambiguity* values contain the number of times the location name is found in *Loc* (*to_ha = 3* in case of *Tripoli*), whereas the *Synonym Ambiguity* values contain the formal name of the location if more than one name is given to the same location (*to_sa = Mumbai* in case of *Bombay*).

*Augmentation*: Using database join operation and while preserving the token order as they appear in the narrative, we augment the token list $To\_V$ with the *Loc* and generate a list of tokens $To\_T$ with values [*to_v, to_cntry, to_ha, to_sa*]. The values of [*to_v, to_ha, to_sa*] are the result of the join operation, while the value of the country [*to_cntry*] is determined based on the ambiguity of the token. If the token is not ambiguous, the value of the country is populated from *Loc*. Otherwise, it is initially set to *null* and will be resolved using in Algorithm 1. Resolving location ambiguity works as follows:

- In case of *Homonym Ambiguity*, we set the country name equal to the value of the country of the city that was last visited or will be visited next, using a contextual sliding window based on the principle of locality, which states that the current place is very likely adjacent to the place that was last visited or will be visited next. We apply this principle when the value of $to\_ha \geq 1$.
- In case of *Synonym Ambiguity* ($to\_sa \neq null$), we set the country of the location token to the country of the formal location.

Algorithm 1 identifies geolocation names, disambiguates the location tokens, assigns the correct country and coordinates to the disambiguated token, preserves the listing order of these geolocation names as they appear in the narrative, and detects whether a specific geolocation name is visited multiple times.

---

**Algorithm 1** TokenDisambiguation

**Require:** $To\_V(to\_v)$ & $Loc(to\_v, to\_cntry, to\_long, to\_lat, to\_ha, to\_sa)$
**Ensure:** $To\_F(to\_v, to\_cntry, to\_long, to\_lat)$
  **Begin**
  $To\_T(to\_v, to\_cntry, to\_ha, to\_sa) \leftarrow To\_V \bowtie Loc$
  // using $To\_T$
  **while** $i \leq n$ **do**
    **if** $to\_ha[i] \geq 1$ **then** // Homonym Ambiguity
      **if** $i = 1$ **then**
        $to\_cntry[i] \leftarrow to\_cntry[i+1]$   // first token in narrative
      **else if** $i = n$ **then**
        $to\_cntry[i] \leftarrow to\_cntry[i-1]$   // last token in narrative
      **else if** $to\_v[i] \in to\_cntry[i-1]$ **then**
        $to\_cntry[i] \leftarrow to\_cntry[i-1]$ // token exist in country of prior
      **else if** $to\_v[i] \in to\_cntry[i+1]$ **then**
        $to\_cntry[i] \leftarrow to\_cntry[i+1]$ // token exist in country of next
      **end if**
    **else if** $to\_sa[i] \neq null$ **then** // Synonym Ambiguity
      $to\_cntry[i] \leftarrow to\_cntry[to\_sa]$
    **end if**
    $To\_T \leftarrow To\_T + (to\_v[i], to\_cntry[i], to\_ha[i], to\_sa[i])$
  **end while**
  $To\_F(to\_v, to\_cntry, to\_long, to\_lat) \leftarrow To\_T \bowtie Loc$
  **End**

---

N2T[+] contains a suite of four tokenization methods:

(1) ST: Significant Entity Tokenization
(2) MWT: Multi-Word Tokenization
(3) ST+Aug+DisAmbig: ST + Geospatial Augmentation + Disambiguation
(4) MWT+Aug+DisAmbig: MWT + Geospatial Augmentation + Disambiguation

**Part of Speech Tagging**: is an optional step that categorizes a token based on its semantics and position in a sentence. Applying NLP POS Tagging techniques to the tokens generated in the Tokenization step creates a list of 2-tuples (token, tag). This step is used to compare the results of only using AI techniques versus combining DS and AI techniques

**Trajectory Charting**: plots the trajectory using the geospatial tokens generated in the tolenization step by depicting the travelled route over time as listed in the narrative.

## 3. EXPERIMENTAL RESULTS
### 3.1 Dataset and Ground Truth

Our text corpora is composed of several human trafficking narratives *(N_1 . . . N_n)*, a multi-word lexicon, and a location DB. The narratives are written by English speaking journalists and acquired from various news agencies and anti-trafficking organizations. They contain ambiguous location names. Their length varies between 605 and 4,212 words per narrative. To measure the performance of *N2T[+]*, we conducted experiments and compared their results against the ground truth. To identify the ground truth, we manually read each of the narrative, extracted the location names, and saved them sequentially in a ground truth structure which was used later on to evaluate *N2T[+]*.

### 3.2 Results

We executed each of the four methods of N2T[+] using each narrative in the text corpus and generated a trajectory. Fig. 2 displays an example trajectory generated by N2T[+] using one migrant trafficking narrative. We compared every trajectory against the ground truth and calculated the performance measures for each method applied.

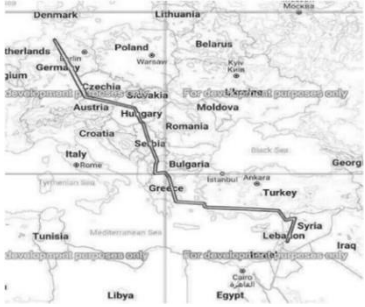

**Figure 2: Trajectory of a Narrative**

Our goal is to extract as many geospatial tokens as possible from text corpora. $ST$ and $MWT$ failed to extract most geospatial tokens; therefore we decided to enhance them by augmenting each of these two methods with location dimension, using context, and applying the location disambiguation approach resulting in methods $ST+Aug+DisAmbig$ and $MWT+Aug+DisAmbig$ respectively. The latter methods do not rely on the geospatial tagging process. They increase the true positive values and reduce false positive outcomes to a value close to zero resulting in higher F1-Score than $ST$ and $MWT$. $MWT+Aug+DisAmbig$ returns the highest F1-Score and Accuracy compared with all other methods because it uses sentence tokenization, Multi-word Lexicon, database joins, context, and the principle of locality to recognize and disambiguate multi-word location tokens as illustrated in Fig. 3.

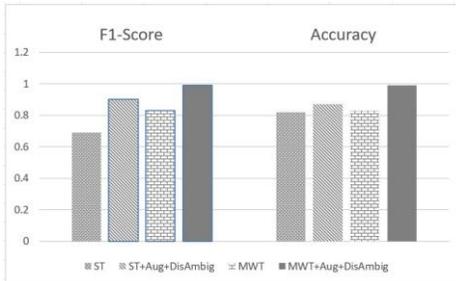

**Figure 3:** Performance Measures of N2T[+] Methods on Trafficking Narratives

We compared our N2T[+] $MWT+Aug+DisAmbig$ to the proposed $ILP$ [2] and $GNER$ [6] and found out that our $MWT+Aug+DisAmbig$ F1Score outperforms $ILP$ by 21% and $GNER$ by 37% as shown in Table 1. Such higher performance is attributed to the thorough data preprocessing, the utilization of NLP libraries, data augmentation, context, and principle of locality.

## 4. CONCLUSION

We presented a system called Narrative to Trajectory (N2T[+]), which extracts trajectories of trafficking routes. N2T[+] uses DS and AI techniques to analyze trafficking narratives, automatically extract relevant location names, disambiguate possible name ambiguities, and plot the trafficking route on a map. We first applied NLP libraries and found out that they lack retrieving some geospatial tokens. We then introduced geospatial dimension augmentation, context, and principle of locality concepts on top of the NLP libraries. In a comparative evaluation we show that our proposed multidimensional approach offers significantly higher geolocation detection and disambiguation than other techniques.

**Table 1: Results of N2T[+] compared with other Methods**

| System | Precision | Recall | F1 Score |
|---|---|---|---|
| $ILP$ | 0.79 | 0.79 | 0.79 |
| $GNER$ | 0.54 | 0.68 | 0.70 |
| $MWT+Aug+DisAmbig$ | 0.98 | 0.91 | 0.96 |

## 5. ACKNOWLEDGEMENTS

This work is funded in part by National Science Foundation (NSF) Award #2118285, "HDR Institute: iHARP Harnessing Data and Model Revolution in the Polar Regions"

## REFERENCES

[1] GeoNames. 2021. GeoNames. https://www.geonames.org/
[2] R. Kapoor, M. Kejriwal, and P. Szekely. 2017. Using Contexts and Constraints for Improved Geotagging of Human Trafficking Webpages.
[3] Saydeh N. Karabatis and Vandana P. Janeja. 2022. Creating Geospatial Trajectories from Text Corpora. In *KDD 2022 Proceedings of the 28th ACM SIGKDD Conference on Knowledge Discovery and Data Mining, Data-driven Humanitarian Mapping, 3rd KDD Workshop*. Association for Computing Machinery (ACM), United States, 1–4. https://kdd-humanitarian-mapping.herokuapp.com/
[4] MEMEX. 2021. DARPA MEMEX. https://www.darpa.mil/program/memex
[5] Magdalena Mis. 2017. Organ trafficking booming in Lebanon as desperate Syrians sell kidneys, eyes. https://www.reuters.com/article/us-mideast-crisis-syria-trafficking/organ-trafficking-booming-in-lebanon-as-desperate-syrians-sell-kidneys-eyes-bbc-idUSKBN17S1V8
[6] Alejandro Molina-Villegas, Victor Muñiz Sanchez, Jean Arreola-Trapala, and Filomeno Alcántara. 2021. Geographic Named Entity Recognition and Disambiguation in Mexican News Using Word Embeddings. *Expert Syst. Appl.* 176, C (aug 2021), 8 pages. https://doi.org/10.1016/j.eswa.2021.114855
[7] C. Nagpal, K. Miller, B. Boecking, and A. Dubrawski. 2017. An Entity Resolution approach to isolate instances of Human Trafficking online.
[8] S. Shahrokh Esfahani, M. J. Cafarella, M. Baran Pouyan, G. DeAngelo, E. Eneva, and A. Fano. 2019. Context-specific language modeling for human trafficking detection from online advertisements.
[9] P. Szekely, C. Knoblock, J. Slepicka, A. Philpot, A. Singh, C. Yin, D. Kapoor, P. Natarajan, D. Marcu, and K. Knight. 2015. Building and Using a Knowledge Graph to Combat Human Trafficking.
[10] E. Tong, A. Zadeh, C. Jones, and L. Morency. 2017. Combating Human Trafficking with Deep Multimodal Models.
[11] UN United Nations. 2015. Transforming our World: The 2030 Agenda for Sustainable Development. https://sdgs.un.org/2030agenda
[12] UN United Nations. 2020. Protocol against the Smuggling of Migrants by Land, Sea and Air, supplementing the United Nations Convention against Transnational Organized Crime. https://www.ohchr.org/en/instruments-mechanisms/instruments/protocol-against-smuggling-migrants-land-sea-and-air
[13] Nick Walsh and et al. 2023. On one of the world's most dangerous migrant routes, a cartel makes millions off the American dream. https://www.cnn.com/2023/04/15/americas/darien-gap-migrants-colombia-panama-whole-story-cmd-intl/index.html
[14] WB World Bank. 2022. Climate and Development in East Asia and Pacific Region. https://www.worldbank.org/en/region/eap/brief/climate-and-development-in-east-asia-and-pacific-region